\documentclass[epjST]{svjour}
\usepackage{graphics}
\usepackage{amsmath}
\usepackage{hyperref}
\usepackage[square,numbers]{natbib}

\begin{document}
\title{Amplitude mode at the superfluid-insulator transition on a random lattice}
\author{Pulloor Kuttanikkad Vishnu \inst{1,2} \and Rajesh Narayanan \inst{1,2} \and Thomas Vojta \inst{3}\fnmsep\thanks{\email{vojtat@mst.edu}}}
\institute{Department of Physics, Indian Institute of Technology Madras, Chennai 600036, India. \and Center for Quantum Information, Communication and Computation (CQuICC), Indian Institute of Technology Madras, Chennai 600036, India. \and Department of Physics, Missouri University of Science and Technology, Rolla, Missouri 65409, USA.}
\abstract{
We study the superfluid-insulator quantum phase transition of interacting bosons by means of large-scale Monte Carlo simulations in the presence of both topological and generic quenched disorders. Recent work has demonstrated that the amplitude mode at this transition broadens and localizes in the presence of dilution disorder, whereas it remains a well-defined delocalized excitation for topological (connectivity) disorder. Here, we analyze the crossover between systems with purely topological disorder and systems with additional bond randomness to disentangle the roles that disorder and interactions play for amplitude mode localization. Specifically, we analyze the scalar susceptibility and its spectral density on both sides of the quantum phase transition. We also discuss the implications of our results for the potential observation of amplitude mode localization in experiments.
} 
\maketitle
%
\section{Introduction}
\label{intro}

As a consequence of the spontaneous breaking of continuous symmetries, two distinct types of collective excitations emerge. They are the gapless Goldstone modes associated with the fluctuations order-parameter angles (phases) and the gapped Higgs modes (also called the amplitude modes) that are associated with the fluctuations of the order-parameter  amplitude \cite{pekker_varma_arcmp_15,shimano_tsuji_arcmp_20}. A variety of techniques have been used to unravel the behavior of these collective modes experimentally: They include Raman scattering-based techniques \cite{measson_gallais_prb_2014},  pump-probe techniques \cite{matsunga_hamada_prl_2013}, and measurements utilizing the Josephson effect in weakly coupled proximate superconductors \cite{carlson_goldman_prl_1975,kadin_goldman_prb_1982}.

A key challenge in observing the collective modes near quantum phase transitions (QPTs) in real quantum materials is posed by impurities, defects, and other random imperfections that are ubiquitous in real materials. It is well known that such quenched disorder leaves a rich imprint on the thermodynamic behavior close to QPTs: For instance, the frozen-in impurities can give rise to new types of universal behavior including infinite-randomness criticality \cite{Fisher92,Fisher95,hoyos_kotabage_prl_07,vojta_kotabage_prb_09} and other kinds of unconventional critical points \cite{AKPR04,AKPR08,vojta_hoyos_jpcm_11}. Disorder effects also include exotic Griffiths singularities \cite{GuoBhattHuse96,RiegerYoung96,VojtaSchmalian05} and smeared transitions \cite{Vojta03a,HoyosVojta08}. In contrast, the impact of disorder on the dynamics of the collective excitations near QPTs has received less attention. Monte Carlo simulations, supported by a quantum mean-field theory, indicate that the amplitude mode near a superfluid-Mott glass transition is localized by quenched disorder \cite{puschmann_crewse_prl_20,crewse_vojta_prb_21,puschmann_getelina_annals_21}. However, the underlying mechanism driving the localization transition has remained unresolved.

Recently, a study based on the Bose-Hubbard model on a random Voronoi-Delaunay (VD) lattice \cite{vishnu_puschmann_prb_2024} (where disorder arises from the random connectivity of the underlying lattice), has shed new light on the localization of the amplitude mode in disordered systems. Ref.~\cite{vishnu_puschmann_prb_2024} reports that the amplitude mode is not localized in the presence of such topological randomness. As it is known that all single-particle states on a 2D random VD lattice are localized \cite{puschmann_cain_epjb_15}, the absence of amplitude localization on the same lattice suggests a many-particle mechanism  rather than a simple quantum-interference mechanism such as Anderson localization. Indeed, the absence of amplitude mode localization at the superfluid-Mott glass transition on a random VD lattice seems to be related to the fact that the critical behavior of the QPT remains in the same universality class as that of the clean problem, consistent with a modified Harris criterion \cite{barghathi_vojta_prl_14}. Ref.~\cite{vishnu_puschmann_prb_2024} also illustrated that the introduction of additional site dilution on the VD lattice leads to a cross-over to a new critical point which lies in the 3D disordered XY universality class and localizes the amplitude mode. This provides further evidence for the amplitude mode localization being moored on the critical behavior of the underlying QPT.

In this paper, we investigate the robustness and universality of the scenario emerging from Refs.\ \cite{puschmann_crewse_prl_20,crewse_vojta_prb_21,vishnu_puschmann_prb_2024} that ties the localization of the amplitude mode to the critical behavior. To this end, we investigate the amplitude mode near the quantum critical point of superfluid-insulator transition on a random VD lattice in the presence of additional bond randomness. Employing a quantum-to-classical mapping of the Bose-Hubbard model onto a (2+1)-dimensional classical XY model, we perform Monte Carlo simulations and analyze the scalar susceptibility via the maximum entropy method (MaxEnt) \cite{jarrell_guernatis_pr_96}.


\section{Model, Simulations, and Data Analysis}

The Bose-Hubbard model on a two-dimensional lattice can be defined by the Hamiltonian
\begin{equation}
    H=\frac{1}{2}\sum_{i}U_{i}(\hat{n}_{i}-\tilde{n})^{2}-\sum_{<ij>}J_{ij}(a_{i}^{\dagger} a_{j} + \textrm{h.c.})~,
    \label{eqn.1}
\end{equation}
where $a_{i}$, $a_{i}^{\dagger}$ are the boson annihilation and creation operators, respectively, at site $i$. Furthermore, $\hat{n}_{i}=a_{i}^{\dagger}a_{i}$ is the number operator, and the constant $\tilde{n}$ is the average filling (which we fix at a large integer value to ensure particle-hole symmetry). The Hubbard interaction at site $i$ is denoted by $U_{i}$ and the nearest-neighbor hopping amplitude between sites $i$ and $j$ is denoted by $J_{ij}$. The translationally invariant version of the Bose-Hubbard model (\ref{eqn.1}) is known to host two phases, namely the Mott insulator (when $U\gg J$) and the superfluid phase (when $J \gg U$), separated by a QPT in the classical 3D XY universality class \cite{WeichmanMukhopadhyay08}.

We are interested in the superfluid-insulator transition in the presence of both topological disorder and bond randomness. We therefore consider the Hamiltonian (\ref{eqn.1}) on a random VD lattice and treat the $U_i$ and $J_{ij}$ as independent random variables. Following Refs.\ \cite{puschmann_crewse_prl_20,crewse_vojta_prb_21,vishnu_puschmann_prb_2024} we perform a quantum-to-classical mapping \cite{WeichmanMukhopadhyay08,wallin_sorensen_prb_94} of the quantum Hamiltonian, arriving at a classical XY model on a layered VD lattice such that the disorder is correlated in the imaginary-time direction,
\begin{equation}
    H_{\rm dis}=-\sum_{\left<i,j\right>,\tau}J^{s}_{ij}\mathbf{S}_{i,\tau}\cdot\mathbf{S}_{j,\tau}-\sum_{i,\tau}J^{t}_{i}\mathbf{S}_{i,\tau}\cdot\mathbf{S}_{i,\tau+1}~.
    \label{eqn.2b}
\end{equation}
Here $\mathbf{S}_{i,\tau}$ is a two-component unit vector, situated at spatial coordinate $i$ and imaginary time coordinate $\tau$.
The interactions in the quantum and classical Hamiltonians are related via $\beta_{c}J^{s}\sim J$ and $\beta_{c}J^{t}\sim 1/U$, where $\beta_{c} = 1/T$ is the inverse temperature of the classical model. This classical temperature does not correspond to the actual temperature of the quantum system, which remains at absolute zero. We can thus tune the system through the transition by varying the classical temperature.

To implement bond randomness, we treat $J^{s}_{ij}$ and $J^{t}_{j}$ as independent random variables, drawn from the power-law distribution \footnote{In principle, universality dictates that the critical behavior of the QPT is independent of the details of the disorder distribution. The power-law distribution is convenient because it becomes very broad on a logarithmic scale for $\Delta \to \infty$, leading to a rapid crossover to the disordered critical behavior. Other $J$-distributions implement weaker bare disorder, leading to a slower crossover and requiring larger system sizes which is numerically expensive.}
\begin{equation}
    P(J)=\frac{1}{\Delta}J^{(-1+\frac{1}{\Delta})}~.
    \label{eq_powerlaw_disorder}
\end{equation}
with $0<J\le 1$. The parameter $\Delta$ can take values between 0 and $\infty$ and serve as a measure of the disorder strength. The clean limit (uniform $J^s$ and $J^{\tau}$) is recovered when $\Delta=0$ whereas the distribution becomes arbitrarily broad (on a logarithmic scale) in the limit $\Delta \to \infty$.

To simulate the classical Hamiltonian (\ref{eqn.2b}), we combine the Metropolis single-spin flip algorithm with the Wolff cluster algorithm. We follow Refs.\ \cite{ballesteros_fernandez_prb_98,vojta_rastko_prb_06,crewse_vojta_prb_21} to reduce the overall computational effort by simulating many disorder configurations while keeping the number of measurement sweeps relatively small.  Specifically, we average over 1000 to 5000 disorder configurations with 500 measurement sweeps each, while limiting equilibration sweeps to 100--200. As is standard practice, we verify the equilibration by comparing simulations starting from both hot and cold configurations.

The order parameter of the transition is the magnetization, given by
\begin{equation}
\mathbf{m}=\frac{1}{N} \sum_{i, \tau} \mathbf{S}_{i, \tau}~,
\end{equation}
where $N$ is the total number of lattice sites. To identify the critical point and to characterize the phase transition, we lay recourse to the Binder cumulant $U_m$ and the order-parameter susceptibility $\chi$, which are given by
\begin{equation}
U_m=\left[1-\frac{\langle | \mathbf{m}|^4\rangle}{3\langle | \mathbf{m}|^2\rangle^2}\right]_{\mathrm{dis}} \quad, \qquad \chi=\frac{N}{T}\left[\langle | \mathbf{m}|^2\rangle-\langle | \mathbf{m}| \rangle^2\right]_{\mathrm{dis}}~.
\end{equation}
Here, $\langle\ldots\rangle$ denotes the thermodynamic (Monte Carlo) average, while the average over disorder realizations is represented by $[\ldots]_{\text {dis}}$.

Quenched randomness breaks the symmetry between the space and imaginary-time directions such that the disordered critical point is no longer described by a Lorentz-invariant theory. Thus, the spatial correlation length $\xi$ and the imaginary-time correlation length $\xi_{\tau}$ differ from each other, implying that the spatial system size $L$ and the imaginary-time size $L_{\tau}$ need to be treated as distinct parameters in the scaling relations. Assuming conventional dynamical scaling, in analogy to Ref.\ \cite{vishnu_puschmann_prb_2024}, the scaling forms of the magnetization, Binder cumulant, and susceptibility read
\begin{equation}
    U_{m}=X_{U}\left(rL^{\frac{1}{\nu}},\frac{L_{\tau}}{L^{z}}\right), \quad  m=L^{-\frac{\beta}{\nu}}X_{m}\left(rL^{\frac{1}{\nu}},\frac{L_{\tau}}{L^{z}}\right), \quad  \chi=L^{\frac{\gamma}{\nu}}X_{\chi}\left(rL^{\frac{1}{\nu}},\frac{L_{\tau}}{L^{z}}\right).
    \label{eqn:scaling1}
\end{equation}
Here $\beta$, $\gamma$ and $\nu$ are the order parameter, susceptibility and correlation length critical exponents, while  $X_{m}$, $X_{\chi}$ and $X_{U}$ are dimensionless scaling functions. We use the anisotropic scaling analysis of the Binder cumulant  \cite{rieger_young_prl_94,sknepnek_vojta_prl_04,vojta_crewse_prb_16}
to identify the transition temperature, and we use the scaling relations (\ref{eqn:scaling1}) to find the critical exponents.

The amplitude mode is an oscillation of the order parameter magnitude in the symmetry-broken (ordered) phase. Since the  variables $\mathbf{S}_{i,\tau}$  in the classical Hamiltonian (\ref{eqn.2b})  have fixed magnitude, we define a fluctuating local order parameter amplitude $\rho$ via coarse-graining as the average of $\mathbf{S}_{i,\tau}$ over a local cluster consisting of site $i$ and all its (spatial) nearest neighbors,
\begin{equation}
    \rho(\mathbf{x}_{i},\tau)=\frac{1}{\kappa_{i}}\left|\mathbf{S}_{i,\tau}+\textstyle\sum_{j}\mathbf{S}_{j,\tau}\right|~.
\end{equation}
Here, $\kappa_{i}$ is the number of sites in the cluster around site $i$. Fluctuations of this amplitude can be obtained from the imaginary-time scalar susceptibility
\begin{equation}
    \chi_{\rho\rho}(\mathbf{x},\tau)=\langle\rho(\mathbf{x},\tau)\rho(0,0)\rangle-\langle\rho(\mathbf{x},\tau)\rangle\langle\rho(0,0)\rangle~.
    \label{eqn:chirhorho}
\end{equation}
The Fourier transformation of (\ref{eqn:chirhorho}) yields the scalar susceptibility $\tilde{\chi}_{\rho\rho}(\mathbf{q},i\omega_{m})$ as a function of wave vector $\mathbf{q}$ and Matsubara frequency $\omega_m$.  The analytic continuation from imaginary (Matsubara) frequencies to real frequencies is a difficult problem; it is sensitive to statistical errors of the numerical data. We perform the analytic continuation by means of the MaxEnt method \cite{jarrell_guernatis_pr_96} (see Refs.\ \cite{puschmann_crewse_prl_20,crewse_vojta_prb_21,vishnu_puschmann_prb_2024} for details of our approach).

\section{Results}
\label{results}

\begin{figure}
\resizebox{\columnwidth}{!}{%
  \includegraphics{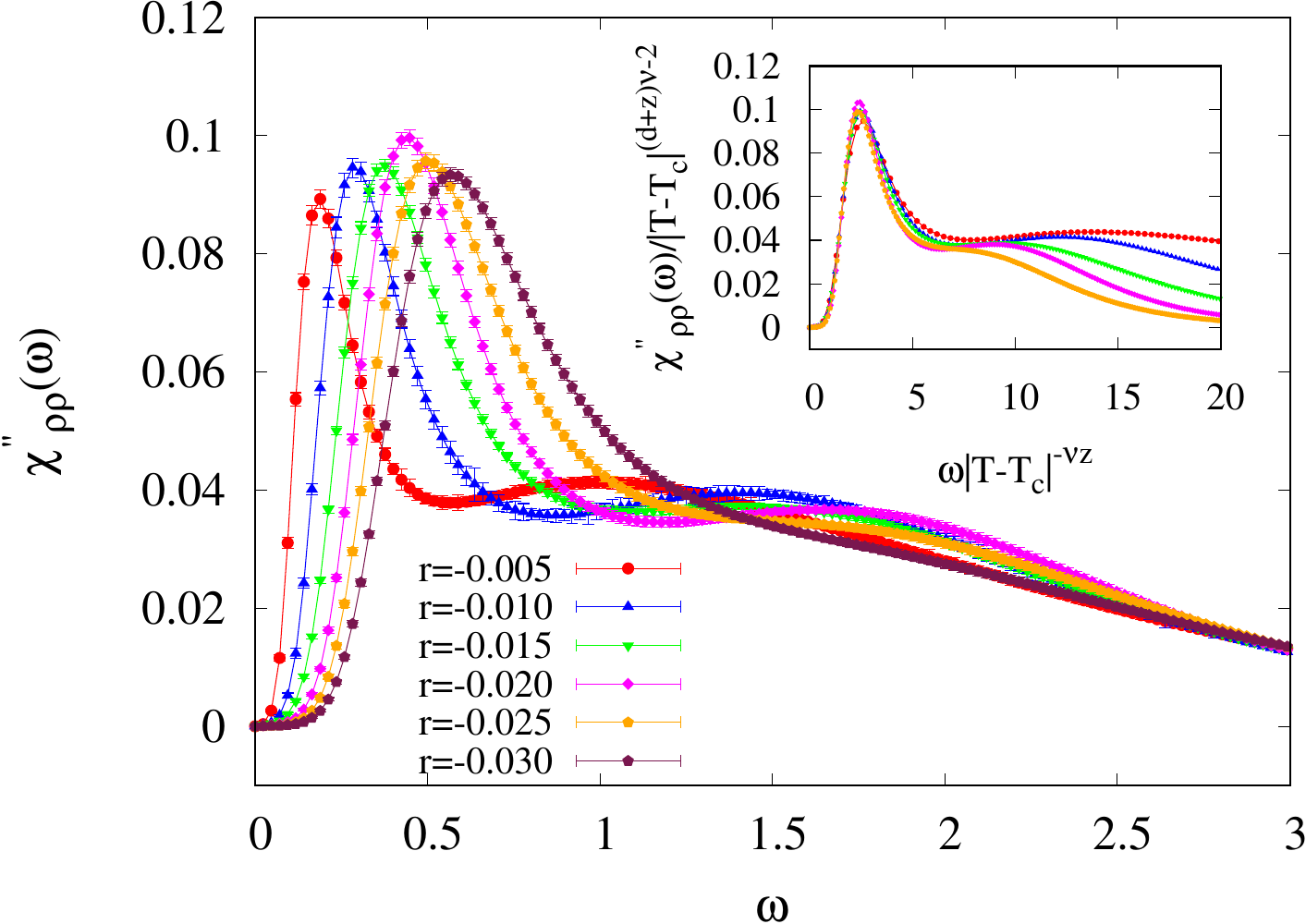}~\includegraphics{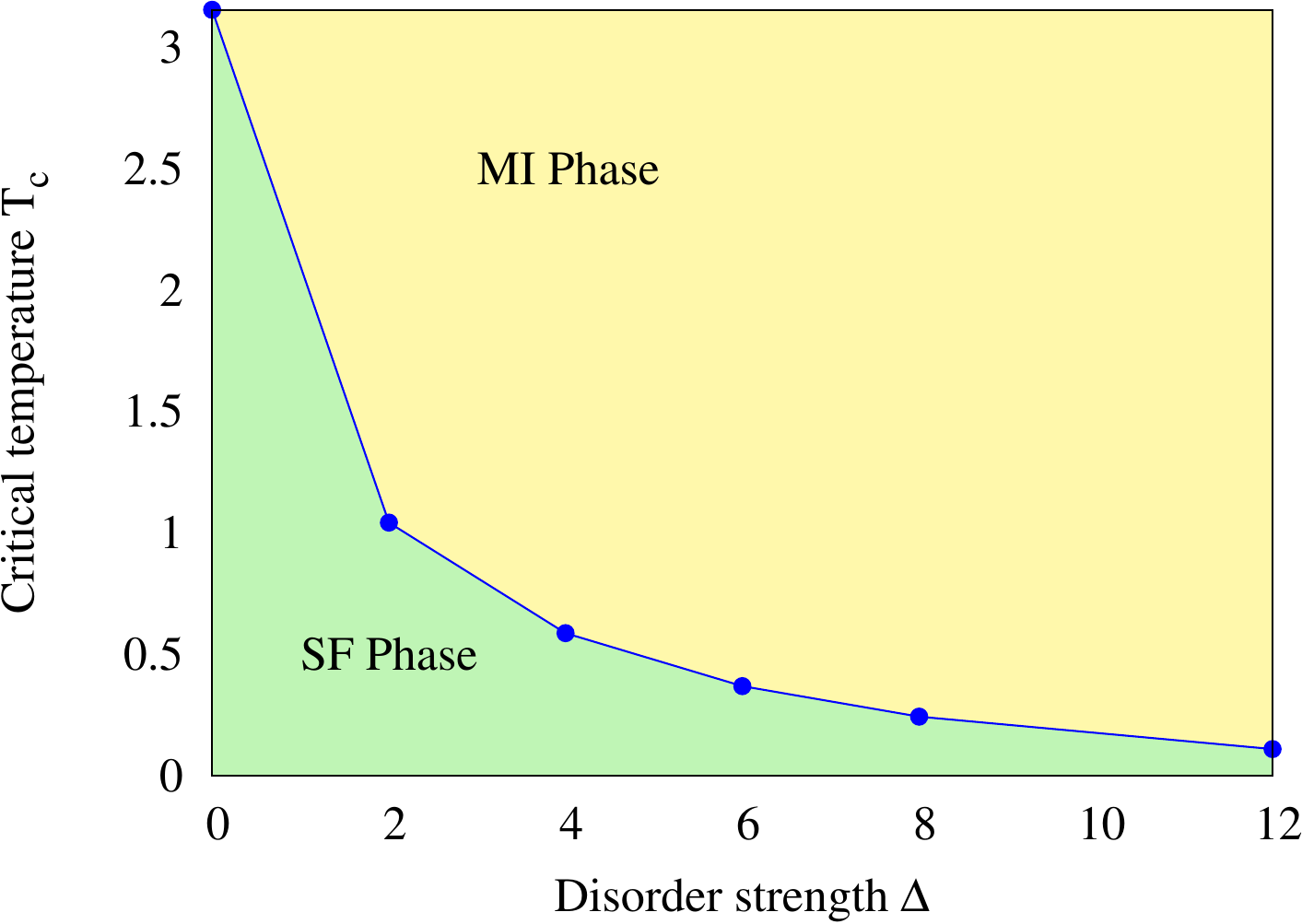}}
    \caption{ (a) Spectral density $\chi^{''}_{\rho\rho}(\mathbf{q}=0,\omega)$ vs.\ real frequency $\omega$ for $\Delta=0$ (pure topological disorder), system sizes $L=L_{\tau}=128$, and different distances $r=T-T_c$ from criticality ($T_c=3.15505(25)$ \cite{vishnu_puschmann_prb_2024}). Inset: Data collapse of the spectral density according to Eq. (\ref{eqn:scaling}). (b) Bond disorder strength -- temperature phase diagram of the system with both bond disorder and topological disorder (i.e., defined on a random VD lattice). }
    \label{fig:voronoi_peaks_and_phase_diagram}
\end{figure}

For later reference, we briefly summarize the properties of the superfluid-insulator transition in the presence of pure topological disorder, i.e., defined on a random VD lattice with uniform interactions \cite{vishnu_puschmann_prb_2024}. Its critical behavior belongs to the clean 3D XY universality class.  Correspondingly, the behavior of the scalar spectral function matches that of the clean case \cite{GazitPodolskyAuerbachArovas13}, as is illustrated in Fig.~\ref{fig:voronoi_peaks_and_phase_diagram}~(a).
The spectral density shows well-defined Higgs peaks; they move to lower energies and become sharper as we approach the critical point, implying that the topological disorder alone cannot localize the amplitude mode. The delocalized behavior of the amplitude mode is further verified by the scaling collapse of the spectral density (shown in the inset of Fig.~\ref{fig:voronoi_peaks_and_phase_diagram}~(a)) according to  \cite{podolsky_sachdev_prb_12,puschmann_crewse_prl_20},
\begin{equation}
    \chi_{\rho\rho}(\mathbf{q},\omega)=\omega^{[(d+z)\nu-2]/(\nu z)}X(\mathbf{q}r^{-\nu},\omega r^{-\nu z})~.
    \label{eqn:scaling}
\end{equation}
We note that previous studies \cite{chen_liu_prl_13,vishnu_puschmann_prb_2024} have observed a Higgs spectral peak not only in the superfluid phase but also in the Mott-insulating phase, provided the system is sufficiently close to criticality, allowing a local order parameter to be defined over a large but finite length scale.

Next, we add bond randomness to the already topologically disordered system by increasing the bond disorder strength $\Delta$ from zero. Following the methodology of Ref.\ \cite{vishnu_puschmann_prb_2024}, we find the transition temperatures as a function of $\Delta$. They are presented in Fig.~\ref{fig:voronoi_peaks_and_phase_diagram}~(b).
The critical temperatures decrease with increasing disorder strength as the weight in the power -law distribution (\ref{eq_powerlaw_disorder}) shifts to lower $J$ with increasing $\Delta$.

To extract the critical exponents of the transition we lay recourse to a finite-size scaling analysis by fitting the system size dependence of various observables right at the critical temperature to the relations  $\chi=aL^{\gamma/\nu}(1+bL^{-\omega})$, $x_{L}=|(d/dT)\ln{|\textbf{m}|}|=aL^{1/\nu}(1+bL^{-\omega})$ and $L_{\tau}^{\rm max}=aL^{z}(1+bL^{-\omega})$. Here, $L_{\tau}^{\rm max}$ is the imaginary-time length at which the Binder cumulant features a maximum as a function of $L$ (see Refs.\cite{rieger_young_prl_94,sknepnek_vojta_prl_04,vojta_crewse_prb_16,vishnu_puschmann_prb_2024}). The fit functions contain corrections to scaling parameterized by the irrelevant exponent $\omega$. As we expect the critical exponents to be universal (independent of the disorder strength $\Delta$), we perform combined fits, for each observable, of the curves for all nonzero $\Delta$ with universal exponent values but $\Delta$-dependent coefficients $a$ and $b$.

\begin{figure}
\resizebox{\columnwidth}{!}{%
  \includegraphics{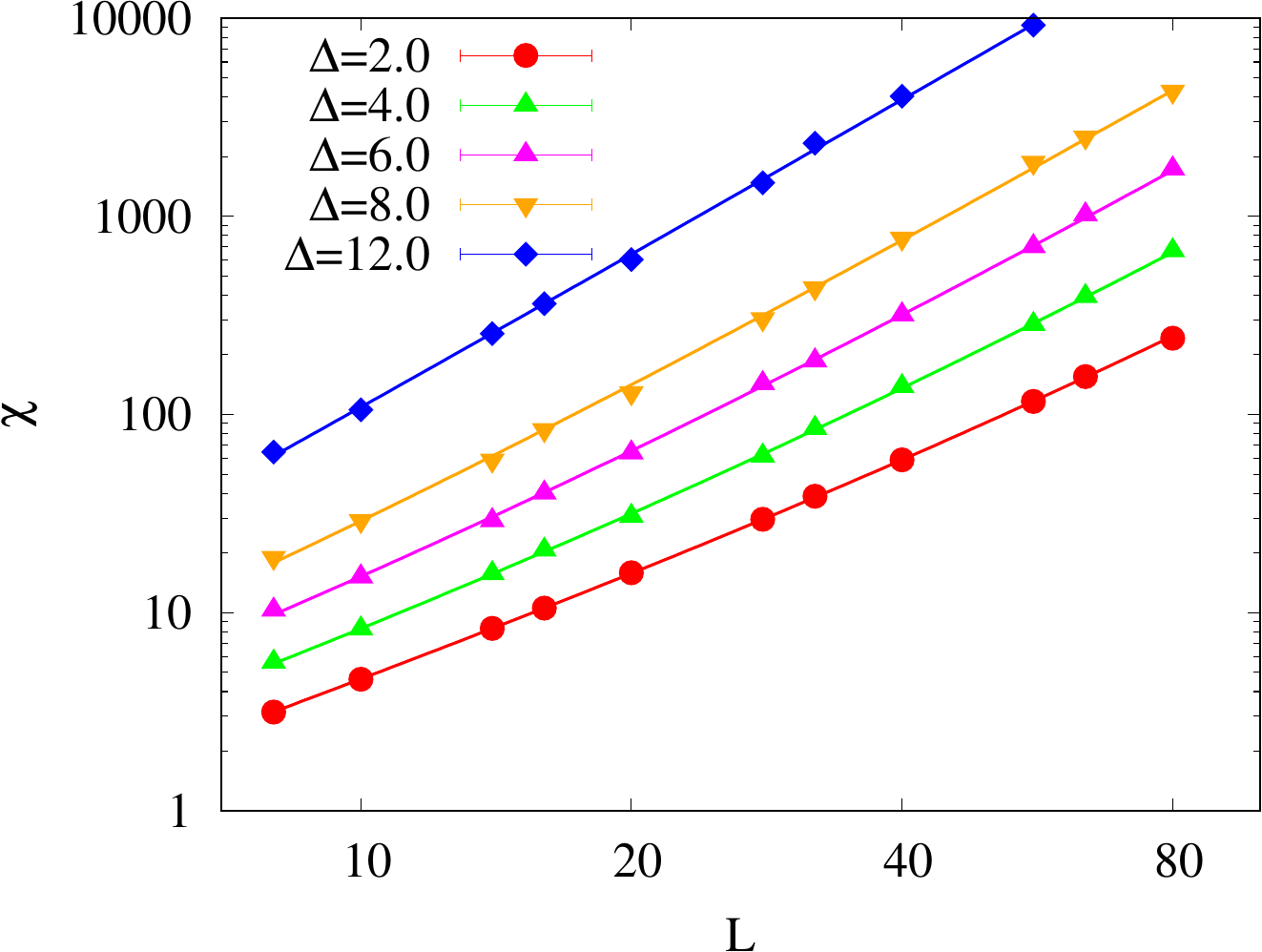}~~~\includegraphics{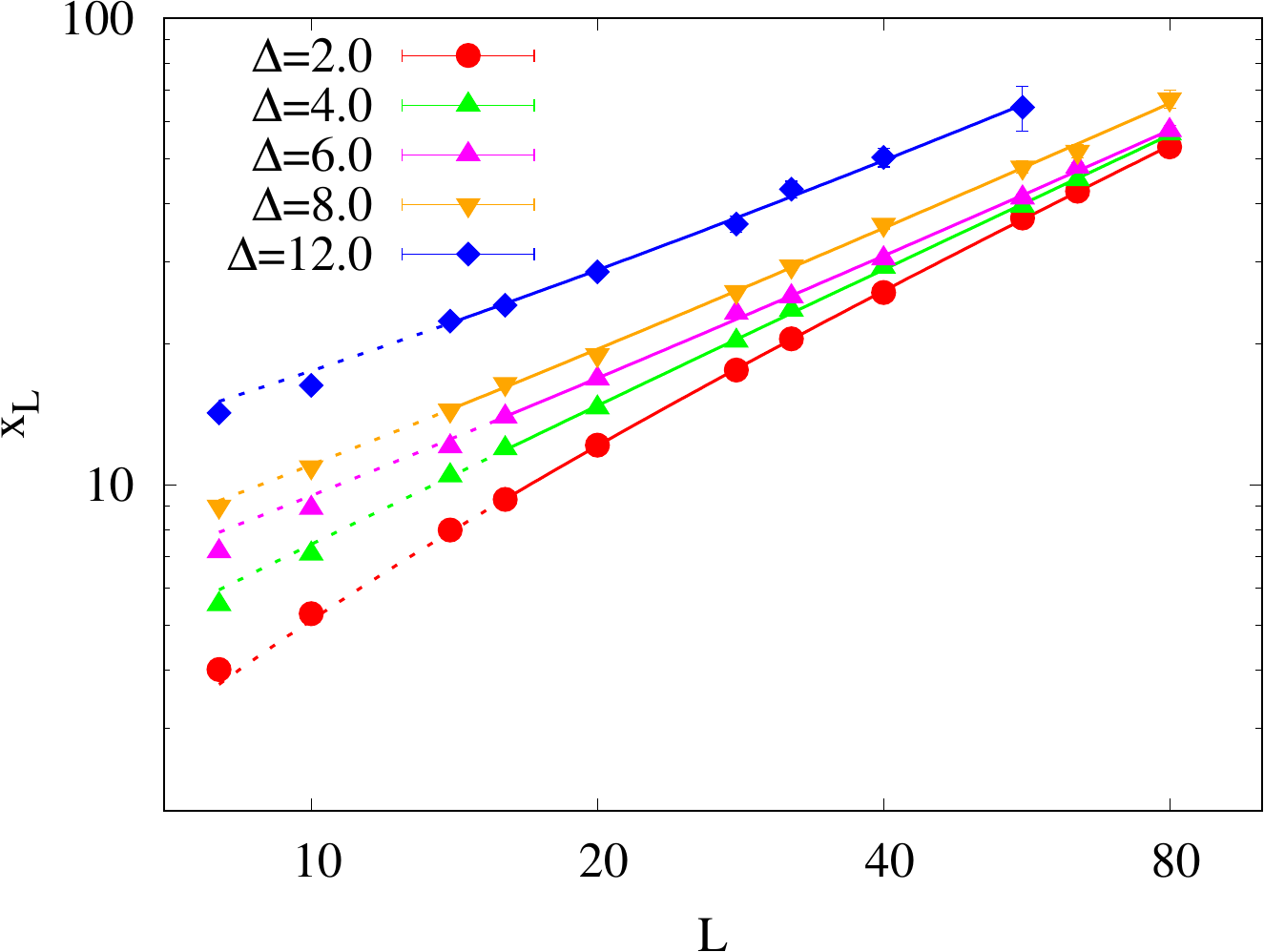}}
    \caption{ Double logarithmic plot of (a) $\chi$ and (b) $x_{L}=|d\ln{|\textbf{m}|}/dT|$ as functions of $L$ at $T_{c}$ for various disorder strengths $\Delta$. The solid lines represent combined fits to (a) $\chi=aL^{\gamma/\nu}(1+bL^{-\omega})$ and (b) $x_{L}=aL^{1/\nu}(1+bL^{-\omega})$ with disorder-independent $\nu$, $\gamma/\nu$ and $\omega$ but disorder-dependent coefficients $a$ and $b$.  The fit for $\chi$ gives $\gamma/\nu=2.6(1)$ and $\omega=1.0(1)$. The fit for $x_{L}$ gives $\nu=1.1(1)$ and $\omega=0.6(1)$. The lines are dotted for smaller system sizes not included in the fits.}
    \label{fig:fits_for_exponents_nu_gamma}
\end{figure}

The analysis of $\chi$ is shown in Fig.~\ref{fig:fits_for_exponents_nu_gamma}(a),  yielding the exponent values $\gamma/\nu=2.6(1)$ and $\omega=1.0(1)$.
Similarly, the combined fit of $x_L$ in Fig.~\ref{fig:fits_for_exponents_nu_gamma}~(b) gives the exponents $\nu=1.1(1)$ and $\omega=0.6(1)$. A similar analysis of $L_{\tau}^{\rm max}$ results in $z=1.5(1)$ and $\omega=0.9(1)$ (not shown). We find that the leading exponent values match with the $(2+1)$-dimensional disordered XY universality class \cite{vojta_crewse_prb_16} within error bars. The values of the irrelevant exponent $\omega$ are slightly different from previous studies
\cite{puschmann_crewse_prl_20,crewse_vojta_prb_21,vishnu_puschmann_prb_2024}. This is possibly related to the fact that our correction-to-scaling term is an effective description of several irrelevant operators whose contributions depend on details of the lattice and disorder distributions.

\begin{figure}
\resizebox{\columnwidth}{!}{%
  \includegraphics{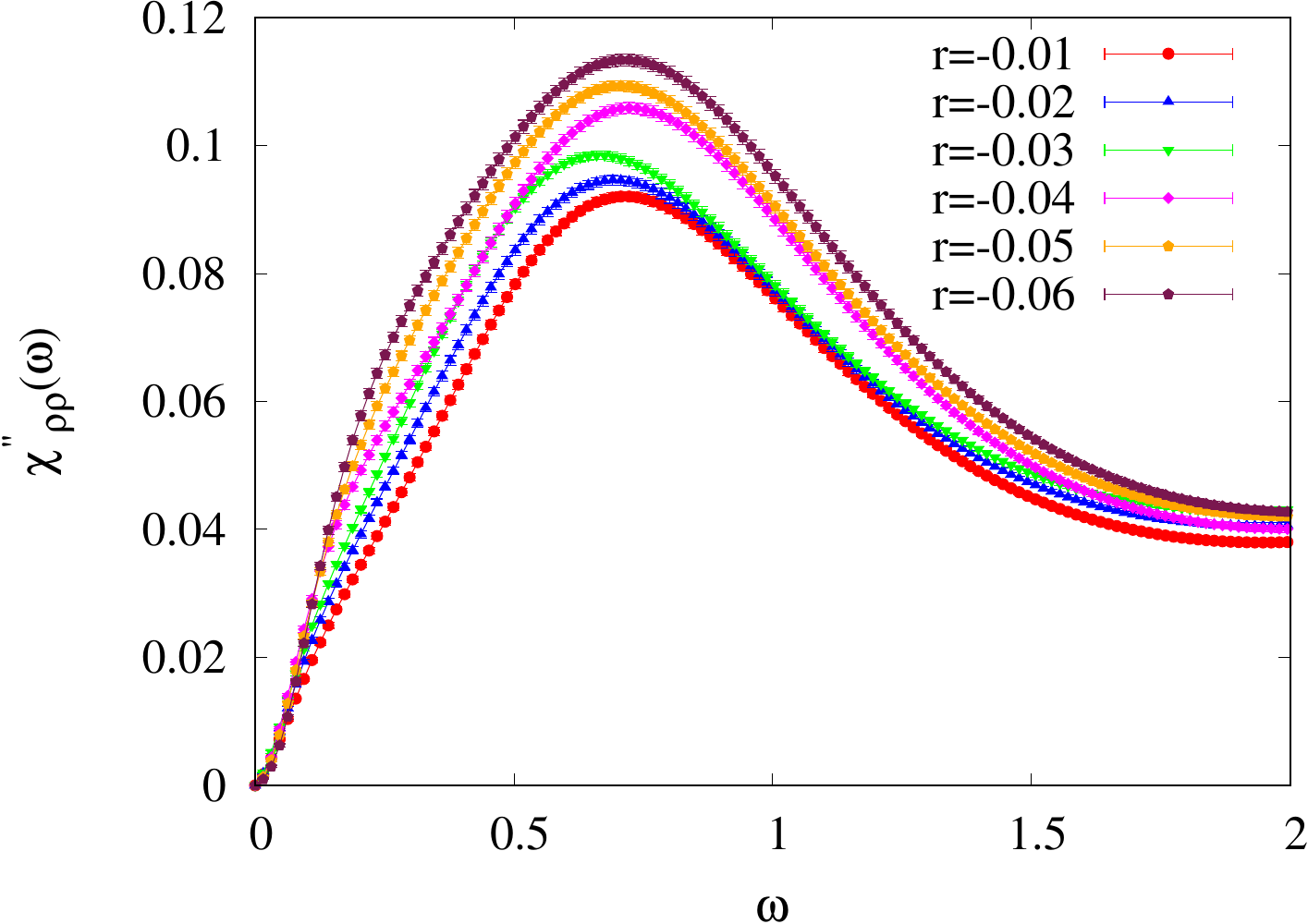}\includegraphics{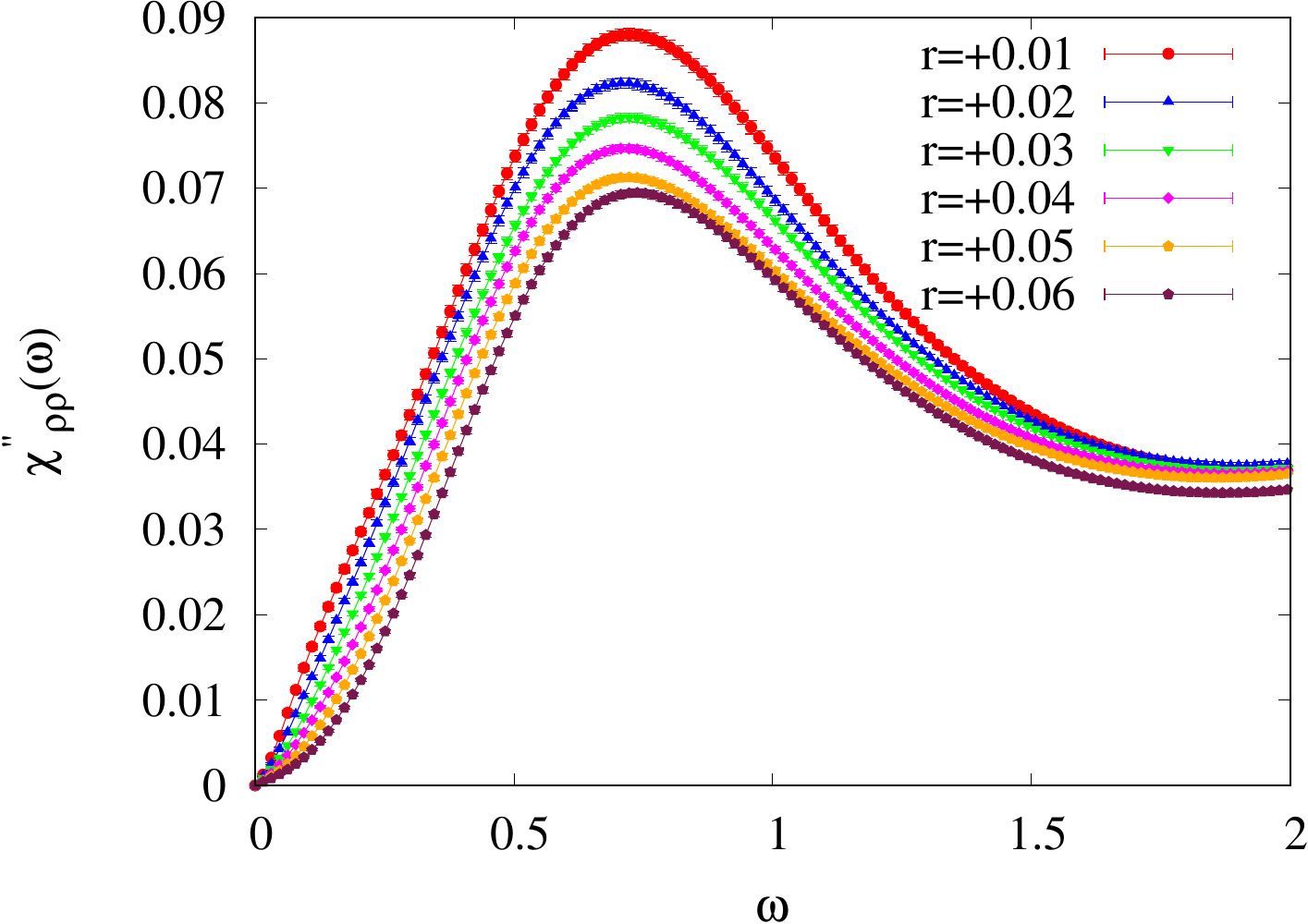}}
    \caption{Spectral density $\chi^{''}_{\rho\rho}(\mathbf{q}=0,\omega)$ as a function of real frequency $\omega$ for different distances from criticality on (a) the superfluid side of the critical point and (b) the Mott side of the critical point for disorder strength $\Delta=4.0$. The size of the lattice used for the simulation is $L=82$ and $L_{\tau}=256$.}
    \label{fig:voronoi_peaks_with_bond_delta_4.0}
\end{figure}

Finally, we discuss the behavior of the amplitude mode in the presence of random-bond disorder on top of the topological disorder. The spectral densities of the amplitude mode for the superfluid side and Mott-insulator side for disorder strength $\Delta=4.0$ are shown in Figs.~\ref{fig:voronoi_peaks_with_bond_delta_4.0}(a) and (b), respectively.
The amplitude spectral densities do not feature Higgs peaks. Instead show only have a broad hump that does not move towards zero energy at the transition. This `localization' behavior is observed in both the superfluid and Mott-insulator phases. We have verified that such a localization behavior of the spectral function is also observed in other disorder strengths that we have studied.

The broadening and suppression of the Higgs peak, as illustrated in Fig.~\ref{fig:voronoi_peaks_with_bond_delta_4.0}, closely parallels the behavior observed at the superfluid--Mott glass transition on a diluted square lattice \cite{puschmann_crewse_prl_20,crewse_vojta_prb_21} and on random-VD lattice with additional vacancy disorder \cite{vishnu_puschmann_prb_2024}. To elucidate this phenomenon, note that the spectral density in the vicinity of a critical point can be decomposed into a singular part and a non-critical regular part as
\begin{equation}
    \chi_{\rho\rho}(\mathbf{q},\omega)=\chi_{\rho\rho}^{\rm reg}(\mathbf{q},\omega) + \omega^{[(d+z)\nu-2]/(\nu z)}X(\mathbf{q}r^{-\nu},\omega r^{-\nu z})~.
    \label{eqn:scaling+regular}
\end{equation}
Here $\chi_{\rho\rho}^{\rm reg}(\mathbf{q},\omega)$ denotes the regular component which is governed by microscopic degrees of freedom, while the second term represents the scaling form (\ref{eqn:scaling}) of the singular component. If the exponents satisfy the condition $[(d+z)\nu-2]>0$,  the singular part of the spectral function is strongly suppressed in the $\omega\rightarrow 0$ limit, leaving the scalar response dominated by the noncritical background. This is exactly the scenario that applies in the bond-disordered case, with $d=2, z=1.5(1)$ and $\nu=1.1(1)$). \footnote{Note that the condition $[(d+z)\nu-2]>0$ always holds if the critical point fulfills Harris' inequality $d\nu >2$ \cite{Harris74,CCFS86}. This suggests that the Higgs peak is generically suppressed at a disordered critical point.} In contrast, at the clean critical point which also governs the transition in a system with purely topological disorder, this exponent combination gives a value close to zero, indicating that the magnitude of the singular component remains nearly constant as the critical point is approached.

\section{Conclusion}
In summary, we have studied the quantum critical behavior and collective excitations near the two-dimensional superfluid-insulator quantum phase transition of interacting bosons on a random VD lattice with additional generic disorder exemplified via random bonds. This analysis was performed by mapping the Bose-Hubbard model (in the limit of large integer filling) onto a classical XY Hamiltonian defined on a (2+1)-dimensional layered VD lattice. The resulting classical model was subsequently investigated using Monte Carlo simulations to determine the universality class of the superfluid-insulator transition. Furthermore, we analyzed the scalar spectral function to investigate the localization properties of the amplitude mode in the presence of both topological disorder and bond randomness.

Our findings unequivocally show that the addition of bond randomness to the topologically disordered VD lattice results in a critical behavior that lies in the disordered (2+1)D XY universality class, analogous to the dilution-disordered case \cite{vishnu_puschmann_prb_2024,crewse_vojta_prb_21,puschmann_crewse_prl_20}. Correspondingly, the bond randomness localizes the amplitude mode near the phase transition. This provides further evidence that the amplitude mode localization does not originate from an Anderson-localization-type mechanism. Instead, it appears to be intrinsically linked to the critical behavior of the quantum phase transition via the scale dimension of the scalar susceptibility.
Our results have wider implications for disordered quantum phase transitions. Utilizing the scaling form in Eq. \ref{eqn:scaling+regular}, one can broadly predict the behavior of the scalar susceptibility at any quantum phase transition governed by a finite-disorder fixed point. An interesting question one could address in the future is the behavior of the amplitude modes at other types of critical points, for example, exotic infinite randomness critical points \cite{Fisher92}.

\bigskip

\noindent \textbf{Acknowledgement:} The simulations were performed on The Mill cluster at Missouri S\&T and the Aqua cluster at IIT Madras. T.V.\ acknowledges support from the NSF under Grants No.\  OAC-1919789 and No.\  PHY-2309135.  P.K.V.\ acknowledges a IIE Travel award by the Global Engagement Office, IIT Madras. R.N.\ and P.K.V.\ also acknowledge funding from the Center for Quantum Information Theory in Matter and Spacetime, IIT Madras, from the Department of Science and Technology, Govt. of India, under Grant No. DST/ICPS/QuST/Theme-3/2019/Q69, and from the Mphasis F1 Foundation via the Centre for Quantum Information, Communication, and Computing (CQuICC).

\bigskip

\noindent \textbf{Data availability statement:} Data sets generated during the current study are available from the corresponding author on reasonable request.

\bibliographystyle{aipnum4-1}

\bibliography{references.bib}

\end{document}